\PassOptionsToPackage{unicode}{hyperref}
\PassOptionsToPackage{hyphens}{url}
\PassOptionsToPackage{dvipsnames,svgnames,x11names}{xcolor}
\documentclass[
]{article}
\usepackage{xcolor}
\usepackage{amsmath,amssymb}
\usepackage{iftex}
\ifPDFTeX
  \usepackage[T1]{fontenc}
  \usepackage[utf8]{inputenc}
  \usepackage{textcomp} % provide euro and other symbols
\else % if luatex or xetex
  \usepackage{unicode-math} % this also loads fontspec
  \defaultfontfeatures{Scale=MatchLowercase}
  \defaultfontfeatures[\rmfamily]{Ligatures=TeX,Scale=1}
\fi
\usepackage{lmodern}
\ifPDFTeX\else
\fi
\IfFileExists{upquote.sty}{\usepackage{upquote}}{}
\IfFileExists{microtype.sty}{% use microtype if available
  \usepackage[]{microtype}
  \UseMicrotypeSet[protrusion]{basicmath} % disable protrusion for tt fonts
}{}
\makeatletter
\@ifundefined{KOMAClassName}{% if non-KOMA class
  \IfFileExists{parskip.sty}{%
    \usepackage{parskip}
  }{% else
    \setlength{\parindent}{0pt}
    \setlength{\parskip}{6pt plus 2pt minus 1pt}}
}{% if KOMA class
  \KOMAoptions{parskip=half}}
\makeatother
\makeatletter
\ifx\paragraph\undefined\else
  \let\oldparagraph\paragraph
  \renewcommand{\paragraph}{
    \@ifstar
      \xxxParagraphStar
      \xxxParagraphNoStar
  }
  \newcommand{\xxxParagraphStar}[1]{\oldparagraph*{#1}\mbox{}}
  \newcommand{\xxxParagraphNoStar}[1]{\oldparagraph{#1}\mbox{}}
\fi
\ifx\subparagraph\undefined\else
  \let\oldsubparagraph\subparagraph
  \renewcommand{\subparagraph}{
    \@ifstar
      \xxxSubParagraphStar
      \xxxSubParagraphNoStar
  }
  \newcommand{\xxxSubParagraphStar}[1]{\oldsubparagraph*{#1}\mbox{}}
  \newcommand{\xxxSubParagraphNoStar}[1]{\oldsubparagraph{#1}\mbox{}}
\fi
\makeatother

\usepackage{longtable,booktabs,array}
\usepackage{calc} % for calculating minipage widths
\usepackage{etoolbox}
\makeatletter
\patchcmd\longtable{\par}{\if@noskipsec\mbox{}\fi\par}{}{}
\makeatother
\IfFileExists{footnotehyper.sty}{\usepackage{footnotehyper}}{\usepackage{footnote}}
\makesavenoteenv{longtable}
\usepackage{graphicx}
\makeatletter
\newsavebox\pandoc@box
\newcommand*\pandocbounded[1]{% scales image to fit in text height/width
  \sbox\pandoc@box{#1}%
  \Gscale@div\@tempa{\textheight}{\dimexpr\ht\pandoc@box+\dp\pandoc@box\relax}%
  \Gscale@div\@tempb{\linewidth}{\wd\pandoc@box}%
  \ifdim\@tempb\p@<\@tempa\p@\let\@tempa\@tempb\fi% select the smaller of both
  \ifdim\@tempa\p@<\p@\scalebox{\@tempa}{\usebox\pandoc@box}%
  \else\usebox{\pandoc@box}%
  \fi%
}
\def\fps@figure{htbp}
\makeatother

\NewDocumentCommand\citeproctext{}{}

\makeatletter
 \let\@cite@ofmt\@firstofone
 \def\@biblabel#1{}
 \def\@cite#1#2{{#1\if@tempswa , #2\fi}}
\makeatother
\newlength{\cslhangindent}
\newlength{\csllabelwidth}
\newenvironment{CSLReferences}[2] % #1 hanging-indent, #2 entry-spacing
 {\begin{list}{}{%
  \setlength{\itemindent}{0pt}
  \setlength{\leftmargin}{0pt}
  \setlength{\parsep}{0pt}
  \ifodd #1
   \setlength{\leftmargin}{\cslhangindent}
   \setlength{\itemindent}{-1\cslhangindent}
  \fi
  \setlength{\itemsep}{#2\baselineskip}}}
 {\end{list}}
\usepackage{calc}

\usepackage{arxiv}
\usepackage{orcidlink}
\usepackage{amsmath}
\usepackage{booktabs}
\usepackage[font=small,labelfont=bf]{caption}
\usepackage{microtype}
\makeatletter
\renewenvironment{CSLReferences}[2]% #1 hanging-indent, #2 entry-spacing
  {\begin{list}{}{%
     \setlength{\topsep}{0pt}%
     \setlength{\partopsep}{0pt}%
     \setlength{\parsep}{0pt}%
     \setlength{\itemsep}{4pt}%
     \setlength{\itemindent}{0pt}%
     \setlength{\leftmargin}{0pt}%
     \ifodd #1
       \setlength{\leftmargin}{\cslhangindent}%
       \setlength{\itemindent}{-1\cslhangindent}%
     \fi}}
  {\end{list}}
\makeatother
\makeatletter
\@ifpackageloaded{caption}{}{\usepackage{caption}}
\AtBeginDocument{%
\ifdefined\contentsname
  \renewcommand*\contentsname{Table of contents}
\else
  \newcommand\contentsname{Table of contents}
\fi
\ifdefined\listfigurename
  \renewcommand*\listfigurename{List of Figures}
\else
  \newcommand\listfigurename{List of Figures}
\fi
\ifdefined\listtablename
  \renewcommand*\listtablename{List of Tables}
\else
  \newcommand\listtablename{List of Tables}
\fi
\ifdefined\figurename
  \renewcommand*\figurename{Fig.}
\else
  \newcommand\figurename{Fig.}
\fi
\ifdefined\tablename
  \renewcommand*\tablename{Table}
\else
  \newcommand\tablename{Table}
\fi
}
\@ifpackageloaded{float}{}{\usepackage{float}}
\floatstyle{ruled}
\@ifundefined{c@chapter}{\newfloat{codelisting}{h}{lop}}{\newfloat{codelisting}{h}{lop}[chapter]}
\floatname{codelisting}{Listing}

\makeatother
\makeatletter
\@ifpackageloaded{caption}{}{\usepackage{caption}}
\@ifpackageloaded{subcaption}{}{\usepackage{subcaption}}
\makeatother
\usepackage{bookmark}
\IfFileExists{xurl.sty}{\usepackage{xurl}}{} % add URL line breaks if available
\hypersetup{
  pdftitle={Continuous biome representations from Earth observation embeddings},
  pdfauthor={Maxwell B. Joseph; Flávia De Souza Mendes; Dieu My T. Nguyen; Camile Sothe; Christopher B. Anderson},
  colorlinks=true,
  linkcolor={blue},
  filecolor={Maroon},
  citecolor={Blue},
  urlcolor={Blue},
  pdfcreator={LaTeX via pandoc}}

\title{Continuous biome representations from Earth observation
embeddings}
\def\asep{\\}
\author{{}Maxwell B.
Joseph~\orcidlink{0000-0002-7745-9990}\textsuperscript{1,*}\asep{}Flávia
De Souza
Mendes~\orcidlink{0000-0002-8918-8462}\textsuperscript{1}\asep{}Dieu My
T.
Nguyen~\orcidlink{0000-0002-5653-4891}\textsuperscript{1}\asep{}Camile
Sothe~\orcidlink{0000-0001-5259-3838}\textsuperscript{1}\asep{}Christopher
B. Anderson~\orcidlink{0000-0001-7392-4368}\textsuperscript{1}\\[0.5em]
\mdseries
\textsuperscript{1} Planet Labs PBC
\\[0.3em]
\mdseries\textsuperscript{*} \href{mailto:max.joseph@planet.com}{max.joseph@planet.com}
}
\date{}
\begin{document}
\maketitle
\begin{abstract}
Biotic communities vary continuously across space, yet biome maps impose
categorical boundaries that compress this variation, particularly at
ecotones where transitional communities are ecologically distinct. Could
Earth observation (EO) foundation models, which encode spectral,
spatial, and temporal information with dense embeddings, convert
discrete biome maps into continuous representations that better capture
ecological variation? Here, we fit a linear classifier on Clay v1.5
satellite image embeddings to predict biome labels from a categorical
map. The softmax output yields a continuous probability vector whose
dimensions correspond to named biome classes. We evaluate this approach
using six Brazilian biomes, 1.3 million embeddings, and 10,015 withheld
forest inventory plots spanning 4,672 plant species. The continuous
biome representation outperforms discrete biome labels for predicting
species occurrence (mean per-species AUC 0.618 vs.~0.570 across 10
spatial cross-validation folds). Decomposing this gain shows that
continuity in the graded probability output, rather than label
reassignment, accounts for the improvement; the pattern holds across all
distances from biome boundaries. The raw 1024-dimensional embedding
remains the strongest predictor we tested (mean AUC 0.646 vs.~0.618),
but the continuous representation recovers most of the embedding's gain
over discrete labels. This simple approach provides a probabilistic
replacement for categorical map labels, preserving their meaning while
encoding graded variation that discrete maps suppress.
\end{abstract}

\section{Introduction}\label{introduction}

Biotic communities on Earth exhibit continuous spatial variation, and
ecologists have long invoked the concept of a biome to classify and
describe this variation (Tansley 1935; Mucina 2019). Map representations
of biomes are typically categorical partitions of space (e.g., forest,
grassland, desert) based on expert knowledge, environmental data, and
information about community composition, and can be further subdivided
into smaller, more detailed ecoregions (Champreux et al. 2024; Olson et
al. 2001; Hargrove and Hoffman 2004; Omernik and Griffith 2014).

Categorical maps encode discontinuities at every boundary, such that
biome categories compress continuous biogeographic variation. Species
respond to environmental gradients individually, and community
boundaries reflect classification choices rather than natural
discontinuities (Gleason 1926; Curtis and McIntosh 1951; Whittaker
1956). Although ecoregion borders capture non-random species turnover at
macroecological scales, the strength of biotic transitions varies across
individual boundaries, such that a graded representation could encode
structure that categorical maps cannot (Smith et al. 2018). These
transitions between communities (ecotones) occur along environmental
gradients and hold distinctive assemblages of their own (Risser 1995).
What is needed, therefore, is a representation that encodes continuous
gradient structure while retaining the interpretive value of named biome
classes.

Methods for producing continuous representations of categorical concepts
have developed in parallel with advances in Earth observation (EO).
Early work developed fuzzy and soft classification methods that assign
pixels partial membership across multiple categories (Fisher and
Pathirana 1990; Wang 1990; Foody 1996). As the volume and variety of EO
data have grown, these approaches have evolved and been applied to a
wide range of problems (Feilhauer et al. 2021). These approaches,
however, tend to require additional data, bespoke workflows, or both.

Recently, EO foundation models have emerged to support continuous Earth
system mapping (Zhu et al. 2026; Brown et al. 2025). Foundation models
trained on satellite imagery produce dense embedding vectors that encode
how the Earth's surface appears across spectral, spatial, and temporal
dimensions (Bommasani et al. 2021; Klemmer et al. 2025). Because
vegetation composition and structure influence surface reflectance, EO
embeddings may also encode ecological gradients (Féret and Boissieu
2020; Rocchini et al. 2022). For example, when paired with field plots,
embeddings can predict species composition and forest structure (Gao et
al. 2025). Plot-level inventories, however, cover a small fraction of
the Earth's land surface. Categorical ecological maps provide named
classes and exist nearly everywhere. If EO embeddings encode information
relevant to ecological variation, they could lift discrete maps into
continuous representations whose dimensions correspond to named
categories.

Here, we derive continuous representations of discrete ecological
concepts from EO foundation model embeddings via linear probing, fitting
a linear classifier to predict map classes from embedding vectors (Alain
and Bengio 2016). We use the probe's softmax output as the continuous
representation. Each dimension corresponds to a category in the original
map, compressing a high-dimensional embedding to an interpretable space
while softening hard categorical boundaries. The method produces a
continuous representation from any categorical map.

To evaluate whether these continuous representations encode ecologically
meaningful structure, we develop a case study using a biome map of
Brazil and \textasciitilde10,000 forest inventory plots. We show that
continuous representations outperform discrete labels for predicting
plant species occurrence, and that the performance gain is attributable
to graded probability outputs rather than label reassignment.

\section{Methods}\label{methods}

\subsection{Data}\label{data}

\begin{figure}

\centering{

\includegraphics[width=1\linewidth,height=\textheight,keepaspectratio]{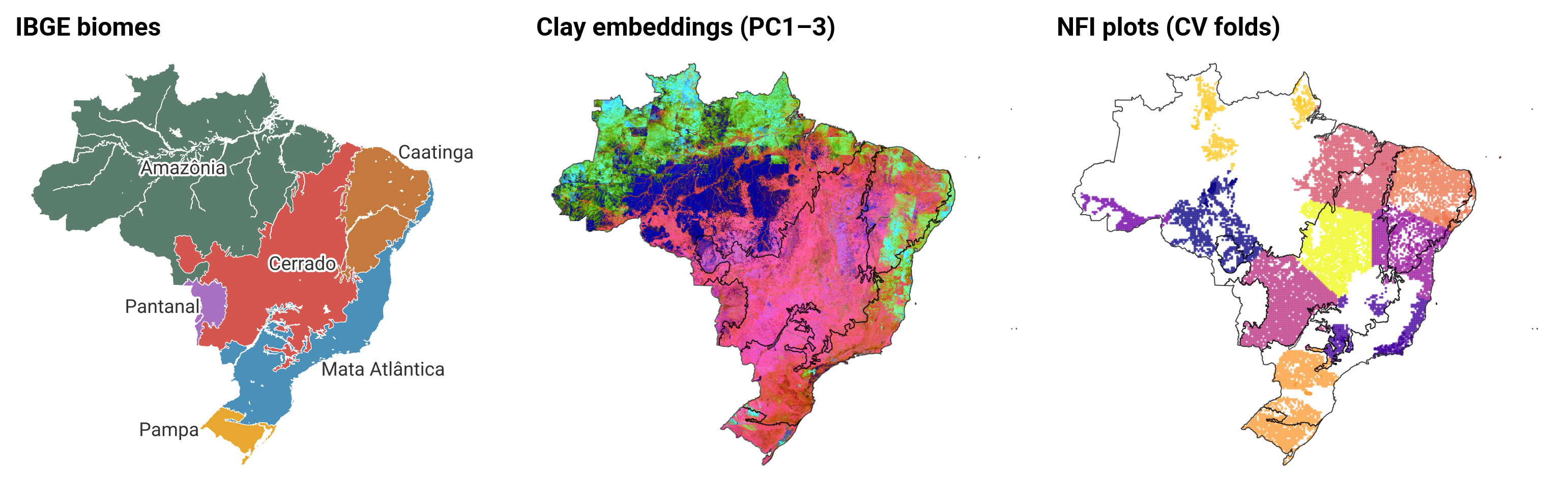}

}

\caption{\label{fig-study-system}Study system overview. Left: IBGE biome
map of Brazil, with six biomes shown as colored polygons. Middle: First
three principal components of Clay v1.5 embeddings rendered as an RGB
composite, with IBGE biome boundaries overlaid. Right: Locations of
10,015 national forest inventory plots, colored by spatial
cross-validation fold assignment, with biome boundaries overlaid.}

\end{figure}%

We develop a case study in Brazil, whose territory spans six Brazilian
Institute of Geography and Statistics (IBGE) biomes
(Fig.~\ref{fig-study-system}, left): Amazônia, Caatinga, Cerrado, Mata
Atlântica, Pampa, and Pantanal (IBGE 2019). From Source Cooperative,
which distributes LGND pre-computed Clay v1.5 embeddings on a
geohash-partitioned grid, we select all tiles intersecting Brazil for
June 2024. Clay v1.5 is a masked autoencoder trained on 10 Sentinel-2
bands spanning visible through shortwave-infrared wavelengths; it
encodes each \(256 \times 256\) pixel tile (approximately 2,560 m on a
side at 10 m ground sampling distance) into a single 1024-dimensional
embedding vector (Fig.~\ref{fig-study-system}, middle). We obtain
approximately 1.3 million such tile embeddings covering Brazil. Biome
labels are assigned for each tile using a majority-area rule within each
tile geometry. The discrete biome map and EO embeddings are the only
inputs to the continuous biome representation.

To evaluate the continuous biome representation, we build a plant
species occurrence dataset from 10,015 Brazilian National Forest
Inventory (NFI) plots, surveyed between 2011 and 2024 (Freitas et al.
2009). NFI plots are distributed unevenly across biomes, with the
majority in Cerrado (3,951) and Amazônia (2,413), followed by Caatinga
(1,735), Mata Atlântica (1,511), and Pampa (405); the Pantanal has no
NFI representation, so the species-level evaluation covers five of the
six IBGE biomes (Fig.~\ref{fig-study-system}, right). The NFI provides
stem-level data from which we compute a binary site-by-species
occurrence matrix spanning 4,672 species. Each NFI plot is assigned the
Clay embedding of its nearest tile centroid.

\subsection{Deriving continuous biome
representations}\label{deriving-continuous-biome-representations}

\begin{figure}

\centering{

\includegraphics[width=0.6\linewidth,height=\textheight,keepaspectratio]{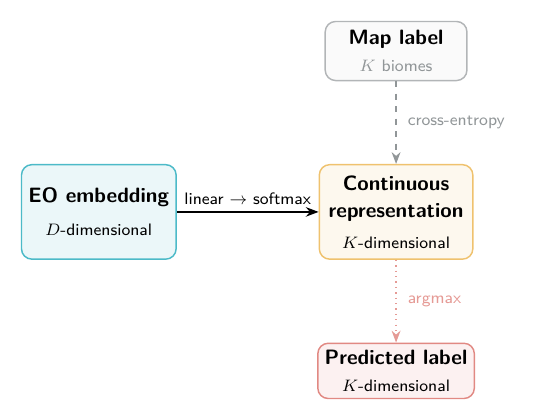}

}

\caption{\label{fig-architecture}A linear layer maps a \(D\)-dimensional
EO embedding to \(K\) logits, and a softmax function produces a
\(K\)-dimensional continuous representation. The model is trained with
cross-entropy loss against discrete biome labels from a categorical map.
Taking the argmax of the continuous representation yields a predicted
label. In this study \(D = 1024\) (Clay v1.5) and \(K = 6\) (IBGE
biomes).}

\end{figure}

We fit a multinomial logistic regression to predict biome labels from EO
embeddings (Fig.~\ref{fig-architecture}). Embedding vectors are first
standardized to zero mean and unit variance per feature, then passed to
a scikit-learn \texttt{LogisticRegression} with default L2
regularization (C = 1) and the \texttt{saga} solver. Because the
categorical map provides labels at every location in the study area, the
classifier is fit once on the full tile set and no spatial extrapolation
is needed. The softmax output of this classifier yields a
\(K\)-dimensional predicted class probability vector (the continuous
representation) whose elements are positive and sum to one. Each element
gives the predicted probability that a tile belongs to a given biome:
when the embedding falls clearly within one class, a single element
dominates the vector; when the satellite-derived signal is consistent
with more than one class, probability mass distributes across multiple
biomes, producing a graded output that discrete labels cannot express.

To characterize uncertainty at each location, we compute the Shannon
entropy of the probability vector, \(H = -\sum_k p_k \ln p_k\), where
\(p_k\) is the predicted probability for biome \(k\). We normalize \(H\)
by its maximum \(H_{\max} = \ln K\) to yield a value in \([0, 1]\).
Normalized entropy, a measure of the ``poorness of a guess'', equals
zero when all probability mass falls on a single biome and one when
probability is uniform across all \(K\) biomes (Wilcox 1973). Locations
with high normalized entropy are those where the continuous
representation departs most from a discrete assignment.

\subsection{Evaluation}\label{evaluation}

We compare three six-dimensional biome representations for predicting
species occurrence, using the raw 1024-dimensional EO embedding as a
reference. The \emph{map label} encodes each location's IBGE biome label
as a one-hot vector. The \emph{predicted label} encodes the classifier's
argmax prediction as a one-hot vector. Substituting the predicted label
for the map label is what we call \emph{label reassignment}. The
\emph{continuous representation} is the full softmax probability vector
defined above. Comparing map label to predicted label isolates the
effect of label reassignment; comparing predicted label to continuous
representation isolates the effect of continuity, i.e., whether graded
probabilities carry information beyond their discrete argmax.

For each representation, we fit per-species L2-regularized logistic
regressions predicting presence/absence under 10-fold spatial
cross-validation, with folds defined by spatially clustering plot
coordinates. We score predictions with per-species AUC, the area under
the receiver operating characteristic curve across a fold's held-out
plots, and include a species in a fold's average only when it has both
presences and absences in the training and held-out sets. We tune the
inverse-regularization strength \(C\) separately for each
representation, searching over a grid from \(10^{-4}\) to \(10^{2}\). We
select \(C\) using an 80/20 validation split within each training set,
maximizing mean validation AUC before refitting on the full training
set. Selected values clustered near \(C = 1\) for the six-dimensional
representations and near \(C = 0.01\) for the raw 1024-dimensional
embedding. Because each model uses a single biome representation (or the
raw embedding) with no additional covariates, absolute AUC values are
expected to be modest. This comparison is designed to isolate the effect
of representation choice, as measured by the per-species difference in
AUC (\(\Delta\)AUC) between representations.

To characterize how the advantage of the continuous representation
varies with proximity to biome transitions and to identify its source,
we stratify held-out plots by distance to the nearest IBGE biome
boundary. For each plot, we determine its assigned biome via spatial
join to IBGE polygons, then compute the distance to the nearest polygon
of any other biome. Plots are binned into distance quartiles. Within
each CV fold and distance bin, we compute per-species AUC for all three
representations and retain only species that are evaluable (i.e., have
both presences and absences) under all three, ensuring that
\(\Delta\)AUC is computed over a matched species set. We then decompose
the mean \(\Delta\)AUC into two components: predicted label minus map
label, and continuous representation minus predicted label.

\section{Results}\label{results}

\begin{figure}

\centering{

\includegraphics[width=1\linewidth,height=\textheight,keepaspectratio]{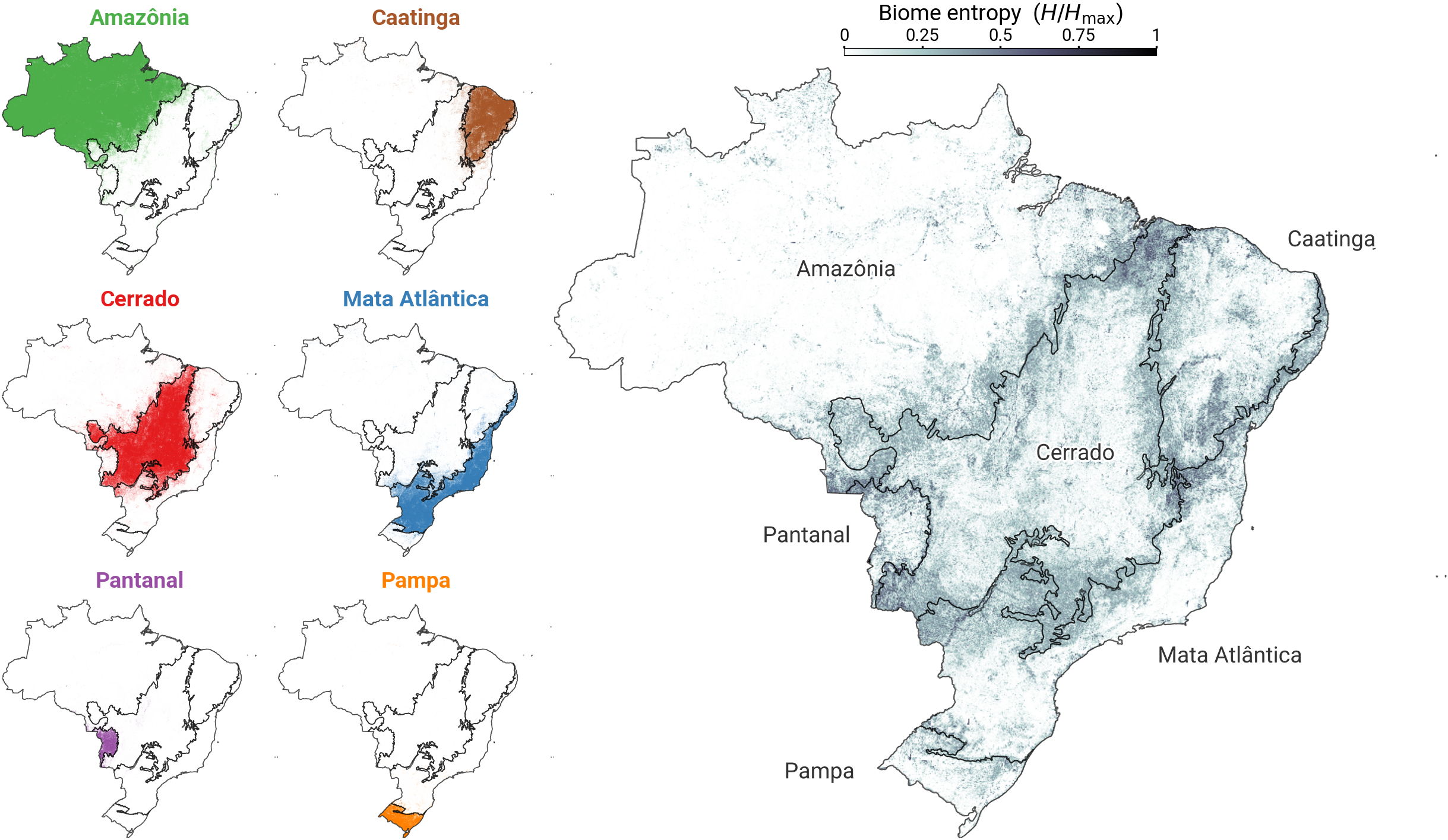}

}

\caption{\label{fig-biome-comparison}Continuous biome representations
derived from Earth observation embeddings. Left: predicted probability
for each of the six IBGE biomes, with color intensity proportional to
probability (saturated = near 1, white = near 0). Right: normalized
Shannon entropy (\(H / H_{\max}\)) of the six-dimensional probability
vector, shown as a sequential shading (light = low entropy, indicating
confident single-biome assignment; dark = high entropy, indicating
probability spread across biomes). IBGE biome boundaries are overlaid on
all panels for reference.}

\end{figure}%

The continuous biome representation preserves the broad spatial
structure of the discrete map, but encodes graded variation that the
discrete map cannot express. In biome interiors, a single class
typically dominates the predicted probability vector, matching the
corresponding IBGE biome; near boundaries, probability mass spreads
across multiple classes, yielding higher entropy
(Fig.~\ref{fig-biome-comparison}). This boundary-associated ambiguity is
unlikely to be driven primarily by mixed tile labels: only 12,274 of
1,309,529 tiles (0.94\%) intersect more than one IBGE biome polygon. The
classifier achieves 93.8\% tile-level accuracy, but the discrete labels
are already available at every location; the quantity of interest is the
full softmax output, a projection of each embedding onto the biome
probability simplex.

\subsection{Continuous representations outperform discrete
labels}\label{continuous-representations-outperform-discrete-labels}

The continuous representation outperforms the map label for predicting
species occurrence in all 10 spatial cross-validation folds
(Table~\ref{tbl-auc-summary}). Mean per-species AUC across folds is
0.618 \(\pm\) 0.052 (mean \(\pm\) s.d.) for the continuous
representation versus 0.570 \(\pm\) 0.055 for the map label. The
predicted label achieves a mean per-species AUC of 0.562 \(\pm\) 0.050,
marginally below the map label, indicating that label reassignment by
itself produces essentially no improvement. The raw 1024-dimensional
embedding is the strongest predictor we evaluated (0.646 \(\pm\) 0.043).
Its advantage over the continuous representation is small but
consistent: it wins 7 of 10 folds, and in the three folds it loses, it
loses by less than 0.01, for a mean advantage of 0.028 AUC across folds.
Using six named biome dimensions rather than 1024 unnamed embedding
dimensions, the continuous biome representation recovers more than half
of the AUC gain.

\begin{longtable}[]{@{}lrl@{}}

\caption{\label{tbl-auc-summary}Mean per-species AUC for predicting
plant species occurrence (mean \(\pm\) s.d. across folds).}

\tabularnewline

\toprule\noalign{}
Predictor & Dimensions & AUC \\
\midrule\noalign{}
\endhead
\bottomrule\noalign{}
\endlastfoot
Map label & 6 & 0.570 \(\pm\) 0.055 \\
Predicted label & 6 & 0.562 \(\pm\) 0.050 \\
Continuous representation & 6 & 0.618 \(\pm\) 0.052 \\
Raw embedding & 1024 & 0.646 \(\pm\) 0.043 \\

\end{longtable}

\subsection{Continuity, not label reassignment, drives the
advantage}\label{continuity-not-label-reassignment-drives-the-advantage}

\begin{figure}

\centering{

\includegraphics[width=1\linewidth,height=\textheight,keepaspectratio]{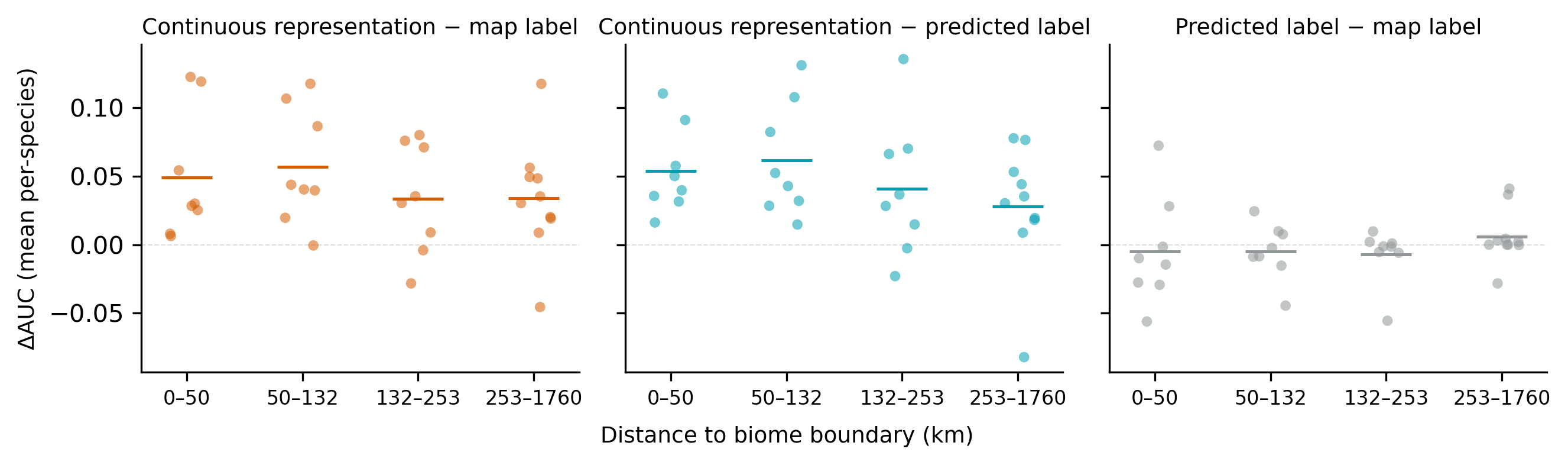}

}

\caption{\label{fig-decomposition}Per-species \(\Delta\)AUC stratified
by distance to the nearest biome boundary. Left: total advantage of the
continuous representation over the map label. Middle and right: additive
decomposition into continuous representation minus predicted label and
predicted label minus map label; the middle and right components sum to
the total shown at left. Each dot is one spatial CV fold's mean
\(\Delta\)AUC across species within a distance quartile; horizontal bars
show the grand mean across folds.}

\end{figure}

The continuous representation outperforms the map label across all
distance quartiles, with the total advantage positive in nearly every
fold--bin combination (Fig.~\ref{fig-decomposition}, left). Decomposing
this advantage into two additive components clarifies its source
(Fig.~\ref{fig-decomposition}, middle and right). The
continuous-minus-predicted component, which isolates the contribution of
graded probability output, is positive in nearly all folds across all
distance bins, with fold-level means of approximately 0.03--0.06
(Fig.~\ref{fig-decomposition}, middle). The predicted-minus-map
component, which isolates the contribution of label reassignment,
scatters around zero across all distance bins
(Fig.~\ref{fig-decomposition}, right). Thus, continuity in the softmax
output, rather than label reassignment, accounts for the gain.

\section{Discussion}\label{discussion}

Biome maps are categorical, but the ecological variation they summarize
is continuous. We introduce a simple way to project EO foundation model
embeddings onto biome space and thereby replace hard labels with a
continuous biome representation. In the Brazil case study, this
continuous representation improved prediction of plant species
occurrence. The decomposition shows that the improvement comes from
continuity in the graded output, not from reassigning biome labels.
Because the primary input is a foundation model embedding of Sentinel-2
data, the graded output captures continuous variation in land surface
conditions observed from space.

The continuous representation is a probability distribution over biome
categories, projecting high-dimensional EO embeddings onto a
low-dimensional named space anchored to an existing map. This
distinguishes it from prior methods which embed discrete ecoregions into
higher dimensional latent spaces (Joseph 2020; Chen and Chiang 2024),
and from unsupervised clustering of foundation model embeddings, which
can discover biome-like clusters that lack predetermined names
(Plekhanova et al. 2026). The raw 1024-dimensional embedding was the
strongest predictor we tested, indicating what the satellite signal
supports under linear probing. The continuous representation trails it
by only 0.028 AUC while compressing 1024 anonymous dimensions into six,
providing a drop-in replacement for categorical map features.

This semantic correspondence matters where biome identity carries legal
force. Under Brazil's Forest Code, legal reserve requirements step from
80\% of property area in Amazônia to 35\% in Cerrado, so that properties
near the Cerrado--Amazônia boundary face a 45-percentage-point
difference in conservation obligation depending on their biome
assignment (Oliveira and Schnaider 2025). Marques et al. (2020) showed
that official maps misclassify ecotonal forests at this transition as
savanna, subjecting high-biomass vegetation to the lower protection
standard and contributing to disproportionate vegetation loss. The
continuous representation flags locations where the EO signal is
consistent with multiple biome categories, but this ambiguity should not
be mistaken for uncertainty in the classification itself: the IBGE map
integrates ground data, expert knowledge, and ancillary sources that
embeddings do not encode. Where high classifier entropy coincides with
stepped legal thresholds, the continuous representation does not resolve
the boundary but instead identifies where resolving it matters most.

One limitation of the present evaluation is temporal mismatch: the Clay
embeddings used here are from June 2024, whereas NFI plots were surveyed
between 2011 and 2024, and the IBGE biome map was published in 2004 and
revised in 2019 (IBGE 2004, 2019). Where land cover has changed between
a plot's survey date and 2024 (e.g., due to deforestation or
degradation), the embedding reflects post-change surface conditions
rather than survey conditions. Because the IBGE biome map is a
biogeographic classification whose boundaries are stable on decadal
timescales, the discrete map label is less susceptible to this source of
noise than the embedding-derived representations, and the measured
advantage of the continuous representation is therefore likely
conservative. Temporally matched embeddings, derived from archival
imagery corresponding to each plot's survey year, would eliminate this
noise source; as foundation model providers expand temporal coverage,
such matching becomes straightforward.

The approach requires only a categorical map and pre-computed
embeddings, and should generalize to other foundation models,
classification systems, and geographies. Beyond direct transfer, several
extensions merit further investigation. On the representation side,
time-indexed embeddings would enable change detection, and richer
pre-training signals (auxiliary losses or additional input modalities)
could expand what a linear probe recovers. On the response side, the
categorical assumption can be relaxed in stages: accommodating nested
classifications (e.g., realm to biome to ecoregion), modeling ranked
categories (e.g., Holdridge life zones) with ordered logistic or probit
regression, and, at the continuous limit, modeling compositional data
(e.g., fractional cover maps) with Dirichlet regression.

The inputs to this approach already exist: pre-trained embeddings cover
most of the terrestrial surface, and categorical biome maps are the
product of decades of biogeographic research. We see the translation
between the two as essential for practical uptake. Methods that bridge
embeddings to named ecological categories should be easy to apply and
should preserve the semantics that practitioners rely on. The approach
developed here meets both requirements. As foundation models improve,
linear projection should recover more ecological structure without
weakening the connection to established ecological concepts.

\section{Acknowledgements}\label{acknowledgements}

The findings and views described herein do not necessarily reflect those
of Planet Labs PBC.

\section*{References}\label{references}
\addcontentsline{toc}{section}{References}

\phantomsection\label{refs}

\end{document}